\documentclass[aps,prl,amssymb,amsmath,twocolumn,floatfix,pre]{revtex4}
\usepackage{graphicx}
\usepackage{amsmath}
\usepackage{color}

\begin{document}
\title{Blume-Capel model analysis with a microcanonical population annealing method}
\author{Vyacheslav Mozolenko$^{1,2}$}
\author{Lev Shchur$^{1, 2}$}
\affiliation{$^1$ Landau Institute for Theoretical Physics, 142432 Chernogolovka, Russia}
\affiliation{$^2$ HSE University, 101000 Moscow, Russia}

\begin{abstract}
We present a modification of the Rose-Machta algorithm (Phys. Rev. E 100 (2019) 063304) and estimate the density of states for a two-dimensional Blume-Capel model, simulating $10^5$  replicas in parallel  for each set of parameters. We perform a finite-size analysis of the specific heat and Binder cumulant, determine the critical temperature along the critical line, and evaluate the critical exponents. The results obtained are in good agreement  with those obtained previously  using various  methods -- Markov Chain Monte Carlo simulation, Wang-Landau simulation, transfer matrix, and series expansion. The simulation results clearly illustrate the typical behavior of specific heat along the critical lines and through the tricritical point.
\end{abstract}

\maketitle

\section{Introduction}
\label{sec:intro} 

The Ising system with anisotropy field~\cite{Blume-1966,Capel-1966}, known as the Blume-Capel (BC) model, is the simplest lattice spin model in which a tricritical point is observed. This is the point on the phase diagram where a smooth line of second order phase transitions suddenly turns into a first order line~\cite{Griffiths-1970,Aharony-1983}. The model has been studied using various analytical and numerical methods, and the status of the study is discussed in detail in the paper~\cite{Butera-2018}.

A new framework for numerical entropy estimation of systems with discrete energy spectrum was recently presented~\cite{Rose-2019}. It is designed to simulate equilibrium systems in a microcanonical ensemble using annealing on the energy ceiling. It has been applied to a first-order phase transition in a two-dimensional 20-state Potts model to study topological transitions in the unstable energy region between two equilibrium states.
One of the microcanonical algorithms proposed in the paper~\cite{Rose-2019}, the microcanonical population annealing algorithm, is an interesting combination of Markov Chain Monte Carlo and population annealing, and is designed to calculate the density of states by estimating entropy. The basic idea is to compute the random transitions of a huge number of replicas and estimate the ratio of replicas that are separated by an energy ceiling as the ceiling goes down the energy spectrum. 

In the original version, the algorithm starts by generating $R$ independent replicas of the system at the maximum energy of the system and moves the replicas down the energy levels using {\em energy ceiling}. The reasons for the modification are several. First, in the general case the choice of the spin configurations with the maximum energy is a difficult task. Second, the number of configurations with maximum energy can be finite. For example, for the two-dimensional Ising model there are only two such configurations, and we would like to have a large number of replicas. Third, the ceiling algorithm of Rose and Machta will drop the systems down the energy spectrum very quickly, and it is impossible to obtain a good enough entropy estimate near the top of the energy spectrum.

 In this paper, we extend the Rose-Machta population annealing approach by introducing {\em energy floor} in addition to {\em energy ceiling}.  The idea of the microcanonical population annealing algorithm (MCPA) is as follows. The randomly generated spin configurations most likely correspond to the most probable energy values centered around the maxima of the density of states (DoS), which is a convex function. Using annealing with energy ceiling, we get the left wing DoS. Using an extension of the procedure with energy ceiling, the procedure with energy floor, we get the right wing DOS.
 
  The accuracy of our approach is additionally verified in a parallel article~\cite{Fadeeva-00}, which compares the results of the Rose-Machta method with the Wang-Landau method~\cite{Wang-Landau,Wang-Landau-PRE} with controlled accuracy~\cite{TM-1} using the Potts model as an example. It turns out that the accuracy of both algorithms, the Wang-Landau algorithm and the ceiling/floor population annealing algorithm, is comparable. The difference between the algorithms is that the Wang-Landau algorithm is a process of random walk in the energy space, and this algorithm is a process of parallel annealing of a sufficiently large population of replicas using ceiling/floor energy annealing.
  
 One of the main problems in simulating using temperature dependent transition probability is the critical slowdown. In the case of second-order phase transitions, the relaxation time increases in a power law with the size of the system in the critical region~\cite{Landau-Binder-book}. In the case of first-order phase transitions, the situation is even worse, since the time scale in the coexistense phase for simulations growth with $L$ as $\exp\left(\Sigma\; L^{d-1}\right)$, where $L$ is the system size, $d$ is the dimension of the system, and $\Sigma$ is the surface tension of the interface ~\cite{VMM-2007}.
  
 A practical feature of the Wang-Landau and energy ceiling algorithms is that the transition probability does not contain any temperature dependence and formally does not have such a critical slowdown. In any case, the characteristic times of the Wang-Landau algorithm, the tunneling time and mixing time, continue to grow according to the power law of the system size in the case of a second-order phase transition~\cite{Shchur-times} and simulating a model with large system size remains problematic. At the same time, it seems that microcanonical algorithms are more preferable for simulating systems with first-order phase transitions~\cite{Janke-2003}.

We choose the Blume-Capel model to check how the modified Rose-Machta framework works for a system with second- and first-order phase transitions, and in the case of a tricritical point, which is very difficult to study numerically.


\section{Model}
\label{sec:model} 

The Blume-Capel model~\cite{Blume-1966,Capel-1966} in the absence of a magnetic field is described by a Hamiltonian 
\begin{equation}
H=-J\sum_{\langle i,j\rangle} \sigma_i \sigma_j +\Delta \sum_i \sigma_i^2,
\label{eq:H}
\end{equation}
\noindent where the spins $\sigma_i$  are located on the sites of a square lattice of linear size $L$ and take one of three values $(-1,0,1)$. Periodic boundary conditions are used, and pairwise interaction of spins  occurs only through the nearest neighbor sites.
The natural variables in the equation~(\ref{eq:H}) are the ferromagnetic coupling constant $J {>} 0$ and crystal field $\Delta$. We will also use the notation $D{=} \Delta {/} J$. The parameter $D$ can be viewed as a disorder parameter, since the model~(\ref{eq:H}) can be mapped to the Ising model with annealed disorder~\cite{Aharony-1983}.

\begin{figure}
	\includegraphics[width=\linewidth]{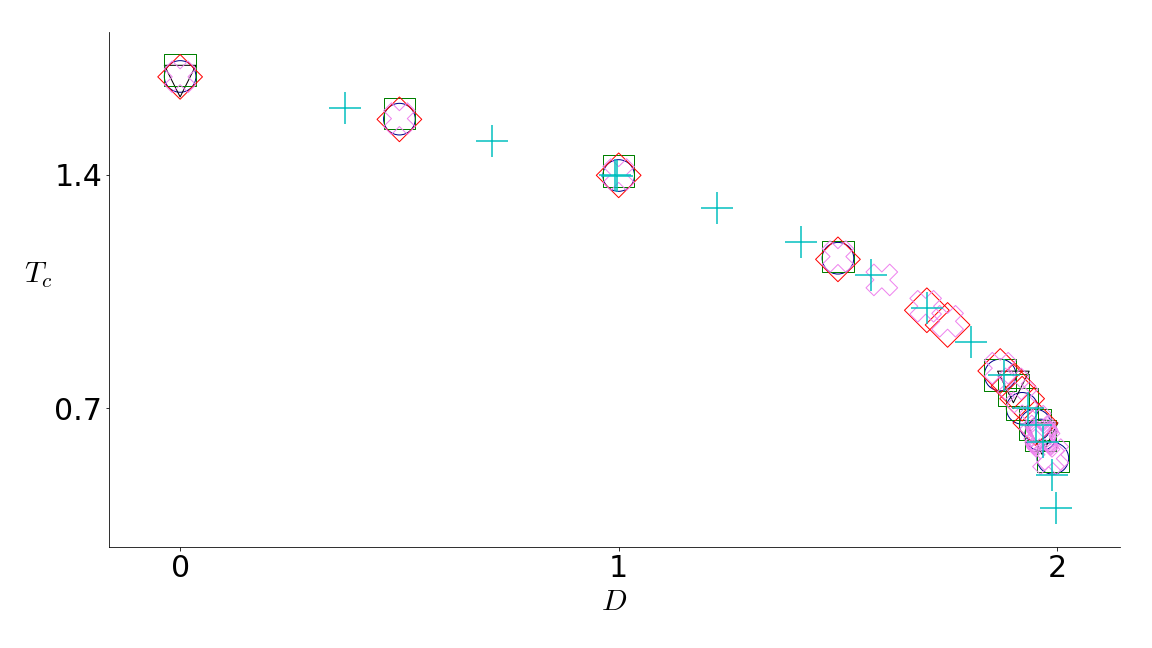}
	\caption{Phase diagram obtained by different methods: transfer-matrix~\cite{Beale-1986} (blue circles), 
Monte-Carlo~\cite{XALP98} (black triangles), Wang-Landau~\cite{SCP06} (green squares), 
high-energy and low-energy expansions~\cite{Enting-1994,Butera-2018} (red diamonds), microcanonical algorithm~\cite{Zierenberg-2017} (cyan plusses), and microcanonical population annealing (current work, violet crosses). The error bars are much smaller than the symbols.}
\label{fig:phase-diagram}
\end{figure}

No exact solutions for the Blume-Capel (BC) model on lattices of dimension $d{>}1$ are known.
The phase diagram obtained by numerical estimation of the continuous phase transition line, the first-order phase transition line and the tricritical point is shown in Fig.~\ref{fig:phase-diagram}. The agreement between the different numerical methods is quite good and the most accepted position of tricritical point is $D_{tr}{\approx} 1.966$ and $T_{tr} {\approx} 0.608$. Our phase diagram estimate is consistent with other methods.

\section{ Microcanonical population annealing}
\label{sec:MCPA}

In this section, we briefly introduce the Rose and Machta ceiling population algorithm presented in Section II of the paper~\cite{Rose-2019}. The authors demonstrate the efficiency of the algorithm in the coexistence phase and are able to capture interesting details of topological transitions in this domain.

In order to apply the ceiling population annealing algorithm to calculating thermodynamic variables, the entire energy spectrum must be covered to be able to estimate entropy. 
So the simulating starts by generating a population of states with maximum energy. Applying the ceiling algorithm~\cite{Rose-2019} will very quickly drive all replicas into the most likely states, and accuracy of density of states (DoS) estimate will be insufficient~\footnote{It should be noted that it is the Wang-Landau probabilities~\cite{Wang-Landau} that drive the system in the Wang-Landau algorithm to the edges of the energy spectrum.}.  
We propose to generate a population with an initially random configuration of spins.

 In this case, the most likely configurations will have an energy corresponding to the maximum of the density of states, which is a convex function with a maximum somewhere in the middle of the energy spectrum (see, for example, Fig.~\ref{fig:entropy}). In this way, we can cover the whole energy spectrum with good statistics and estimate the DoS by combining the results of the ceiling and floor simulations using an appropriate procedure. 
We present the ceiling and floor processes in a unified description.

\subsection{Rose-Machta ceiling procedure}

Rose and Machta's approach to simulating equilibrium systems in a microcannonical ensemble does not relax with temperature; instead, the independent variable of the algorithm is energy. The MCMC procedure consists of a single spin-flip algorithm. The moves occur in  configuration space, and the probability of transition from the state $\alpha$ to the state $\omega$ with energy $E_\omega$ is given by 
\begin{equation}
P_{ceiling}(\alpha\rightarrow \omega) = \left\{
\begin{array}{lcl}
1 & {\rm if} & E_\omega \le E_c  \\ 
0 & {\rm if} & E_\omega > E_c
\end{array}
\right.,
\label{eq:prob-cooling}
\end{equation}
where $E_c$ is the value of ceiling energy, the {\em cooling} energy value. The elementary MCMC step consists with $N$ updating of randomly chosen spins with $N$ is the number of spins. Another parameter of the algorithm is the number of elementary MCMC steps $n_s(E)$.

The algorithm satisfies detailed balance, but is not ergodic for all ceiling energies $E_c$, especially near a ground state consisting of more than one ordered state. Ergodicity can be ensured by simulating a sufficiently large number of parallel replicas~\cite{Rose-2019}. The success of the method was demonstrated by the example of a strong first-order phase transition in the Potts model with 20 states~\cite{Rose-2019}. The method is in a sense a mixture of three algorithms: the simulated annealing algorithm~\cite{Gelatt-1983},  the Wang-Landau algorithm~\cite{Wang-Landau,Wang-Landau-PRE}, and the populations annealing algorithm~\cite{Machta-2010,PA-Review}.

\subsection{Floor microcanonical procedure}

Extension of this idea is annealing using the floor instead of ceiling. The moves occur with the probability of transition from the state $\alpha$ to the state $\omega$ with energy $E_\omega$ is given by 
\begin{equation}
P_{floor}(\alpha\rightarrow \omega) = \left\{
\begin{array}{lcl}
1 & {\rm if} & E_\omega \ge E_f  \\ 
0 & {\rm if} & E_\omega < E_f
\end{array}
\right.,
\label{eq:prob-heating}
\end{equation}
where $E_f$ is the value of floor energy, the {\em heating} energy value. 

\begin{figure}[h!]
	\includegraphics[width=\linewidth]{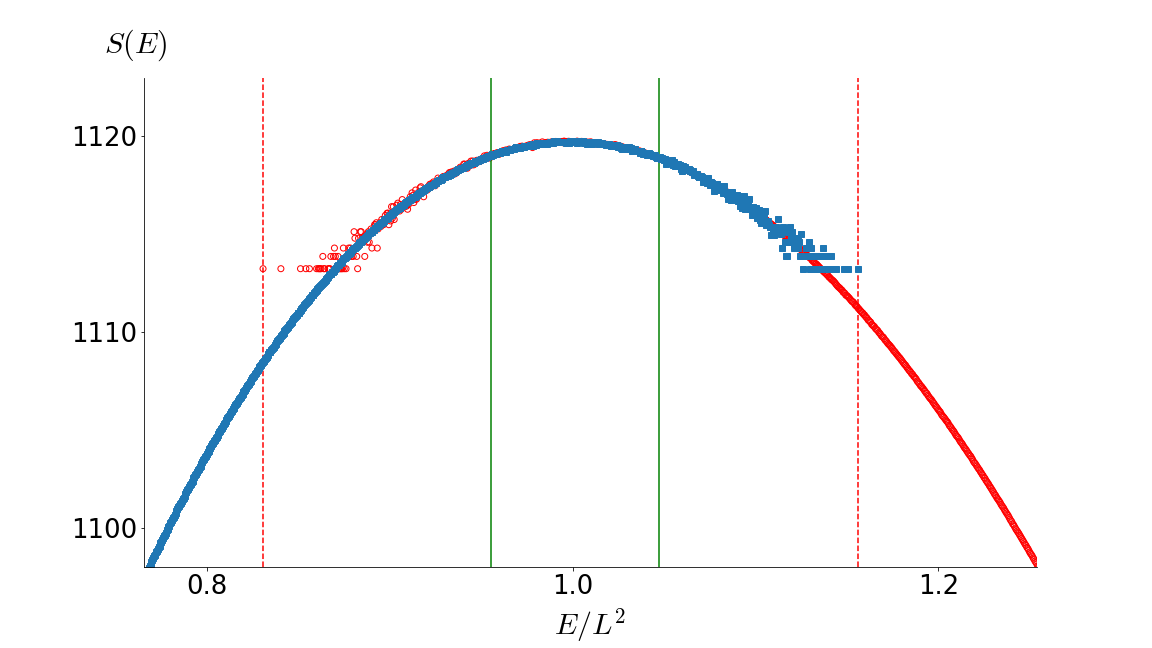}
		\caption{Example entropy estimate for Blume-Capel model with $D=1.5$ and linear system size $L=32$. The blue solid squares are calculated using the ceiling algorithm, the red open circles are calculated using the floor algorithm, the right dashed red vertical line marks the rightmost value of $S^c(E)$, the left dashed red vertical line marks the leftmost value of  $S^f(E)$. The inner green solid vertical lines mark the stitching region. Details are given in the text.}
	  \label{fig:SE-BC-32}
\end{figure}

\subsection{Microcanonical population annealing algorithm}

Combining the ceiling and floor procedures, we get the following algorithm. We do not apply any annealing protocols, but instead perform lowering and raising of energy levels that are calculated dynamically.

{\it Initialization --} Generate $R$ copies of a system with a random spin configuration, most of which are located near the maximum of the energy probability distribution $g(E)$, called the density of states (DoS). Set the initial value of the ceiling $E_c(0)$ to the maximum energy in the ensemble of replicas, and the initial value of the floor $E_f(0)$ to the minimum energy.

{\it Elementary step $i$ of the algorithm --}

1. Perform the $n_s(E_c(i))$ ($n_s(E_f(i))$) MCMC steps, thereby creating new configurations $R$, which represent a pool of configurations.

2c. {\it Ceiling step --} Set the next value of the ceiling $E_c(i)$ to the nearest lower energy level in the pool configurations.

2f. {\it Floor step --} Set the next value of the floor $E_f(i)$ to the nearest higher energy level in the pool configurations.
 
3. Count the number of replicas in the pool $R'$ with energy $E_c(i)$ ($E_f(i)$), and calculate the culling fraction  $\epsilon(E_c(i)){=}R'/R$ or  $\epsilon(E_f(i)){=}R'/R$. Filter these $R'$ configurations from the pool of configurations. 

5.  Randomly select configurations from the pool until the number of replicas equal $R$.

6c. Go to step 1 until the lowest energy for the ceiling is reached. 

6f. Go to step 1 until the highest energy for the floor is reached. 

\subsection{Stitching the parts of entropy together}

To estimate the extensive part of entropy~\cite{Rose-2019}, culling fractions are used for the ceiling and floor
\begin{eqnarray}
S^c(E) &=& \ln(\epsilon(E)) + \sum_{E' > E} \ln(1 - \epsilon(E')), \\
S^f(E) &=& \ln(\epsilon(E)) + \sum_{E' < E} \ln(1 - \epsilon(E')).
\label{eq:S_calc}
\end{eqnarray}
Entropy allows us to add arbitrary constants, which we denote as $S_0^c$ and $S_0^f$, the entropy constants for the ceiling and floor, respectively.

As can be seen from the simulation example in Fig.~\ref{fig:SE-BC-32}, both cooling and heating only cover one wing of the entire energy spectrum. The intersection spanned by both runs is near the entropy maximum, where the random initial replicas are probably located. We obtain entropy over the entire energy range by stitching together the cooling and heating wings in the overlapping region.

Stitching is a somewhat arbitrary procedure that is not sensitive to selection details. We perform it as follows:

1. Select the intersection area bounded by the outer red vertical lines from the leftmost point of the floor wing to the rightmost point of the ceiling wing, see Fig.~\ref{fig:SE-BC-32}.

2. The ends of the ceiling and floor wings are somewhat scattered, so we cut off the outer {\em thirds} areas, leaving us with the area bounded by the inner green vertical lines, which we denote as $E_{left}$ and $E_{right}$.

3. Calculate the mean difference
$$\Delta S(E)=\sum_{E\in [E_{left},E_{right}]}\frac{S^c(E) - S^f(E)}{N_{stitch}},$$ where $N_{stitch}{=} {\mathrm{number\;of\; levels\; in\;}  [E_{left},E_{right}]}$, i.e. for all energies within the green lines in Fig.~\ref{fig:SE-BC-32}, for energy levels coming from $E_{left}$ and $E_{right}$. This allows us to write the stitched $S(E)$ in the form 
\begin{equation}
    S(E) =  S_0 + \left\{
    \begin{array}{lll}
        S^c(E)                             & {\rm if} E < E_{left} \\ 
        S^f(E) + \Delta S(E)                  & {\rm if} E > E_{right} \\
        \frac{S^c(E) + S^f(E) + \Delta S(E)} {2} & {\rm else}.
    \end{array}
    \right.
    \label{eq:S_diff}
\end{equation}

4.  The last free constant $S_0$  can be fixed by counting the number of all states in the system, which is $3^{L^2}$ in our case of the Blume-Capel model on a square lattice with $L^2$ sites
\begin{equation}
    \sum_{\{E\}} e^{S(E)} = 3^{L^2}.
    \label{eq:S_0}
\end{equation}

It should be noted that step 3 is necessary to stitch together the left and right entropy wings, since in the general they should not coincide. The last step 4 consists in normalizing the DoS by the number of possible states, which should lead to the correct entropy values.

\subsection{Estimation of the thermodynamic observables}

An estimate of the canonical partition function is given as a function of temperature $T$ measured in energy units 
\begin{equation}
Z(T) = \sum_E  e^{- E/T  + S(E)}
\label{eq:Z}
\end{equation}
up to some constant multiplier. This multiplier canceled when computing the estimates of the canonical averages, so for simplicity we simply omit it and consider only the extensive part of entropy.

The estimates of average internal energy $\langle E(T)\rangle$ and specific heat $\langle C(T)\rangle$ at temperature $T$ are calculated using the following expressions
\begin{eqnarray}
\langle E(T)\rangle  &=& \frac{\sum_E E\; e^{- E/T  + S(E)}}{Z(T)}, \label{eq:E} \\
\langle E^2(T)\rangle  &=& \frac{\sum_E E^2\; e^{- E/T  + S(E)}}{Z(T)}, \label{eq:E2}\\
C(T) &=& \frac{\langle E^2(T)\rangle - \langle E(T)\rangle^2}{T^2}. \label{eq:C}
\label{eq:PET}
\end{eqnarray}

\section{Algorithm realization}
\label{sec:details}

The implementation of the algorithm is based on a modification~\cite{Rose-2019} of the accelerated population annealing algorithm for GPU~\cite{GPU-PA-2017}. The simulation were performed on an NVIDIA V100 GPU with a typical replica number  $R{=}2^{17}{=}131072$.

The number $R$ of initial and independent replicas of the system~(\ref{eq:H}) is generated with random spin configurations. Thus, these will be the most probable configurations with the most probable energies. We use the cuRAND package~\cite{cuRAND} with the Philox random number generator from the CUDA SL package, which allows us to have independent sequences of pseudorandom numbers. The largest linear lattice size in our research is $L{=}96$, and each ceiling/floor simulation in one replica uses about $2\;10^5$ random numbers per algorithm step. The total number of steps is equal to the number of energy levels, which is about $10^6$. The total number of random numbers per run of one replica is about $2^{40}$, which is less than the length of Philox stream $2^{64}$. This concludes the discussion that parallel replica simulations are random and uncorrelated. 

\begin{figure}[h!]
	\includegraphics[width=\linewidth]{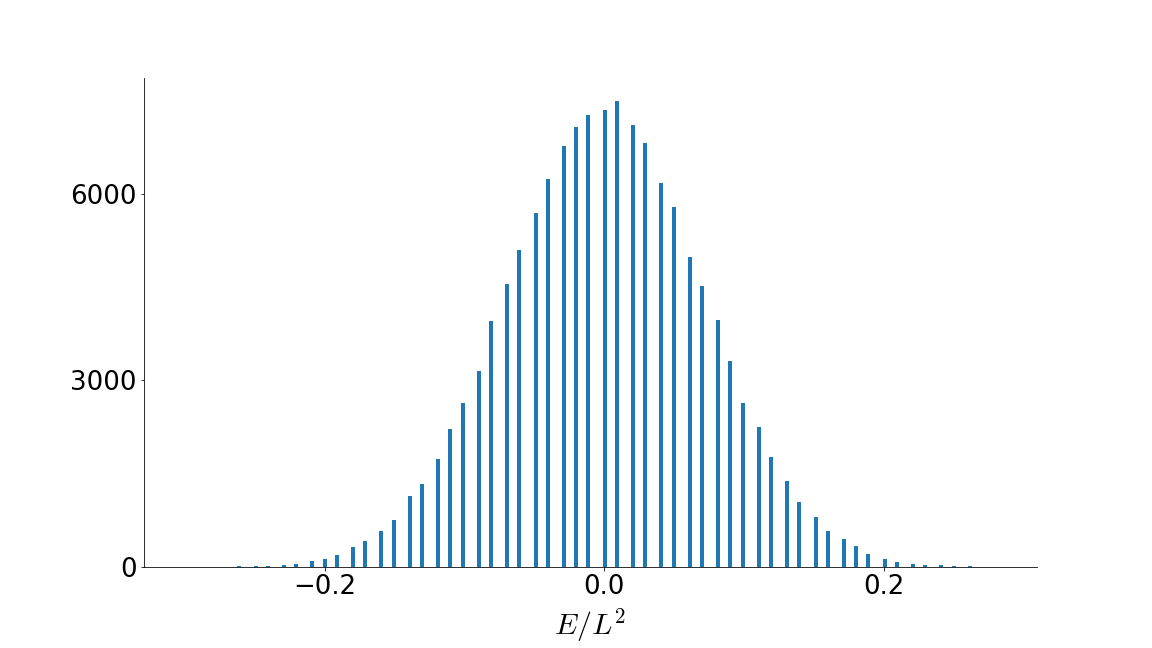}
		\caption{Number of replicas with energy $E$: initial random configuration of 2D Ising model with square lattice size $L=20$. }
	  \label{fig:initial}
\end{figure}

We validate our algorithm by computing the DoS of the 2D Ising model and comparing them well with the corresponding exact solution~\cite{Beale-1996} using $n_s(E){=}10$ steps. Figure~\ref{fig:initial} shows the distribution of the number of initial replicas of 2D Ising model with $L=20$ as function of energy, which is centered around zero. 

\begin{figure}[h!]
	\includegraphics[width=\linewidth]{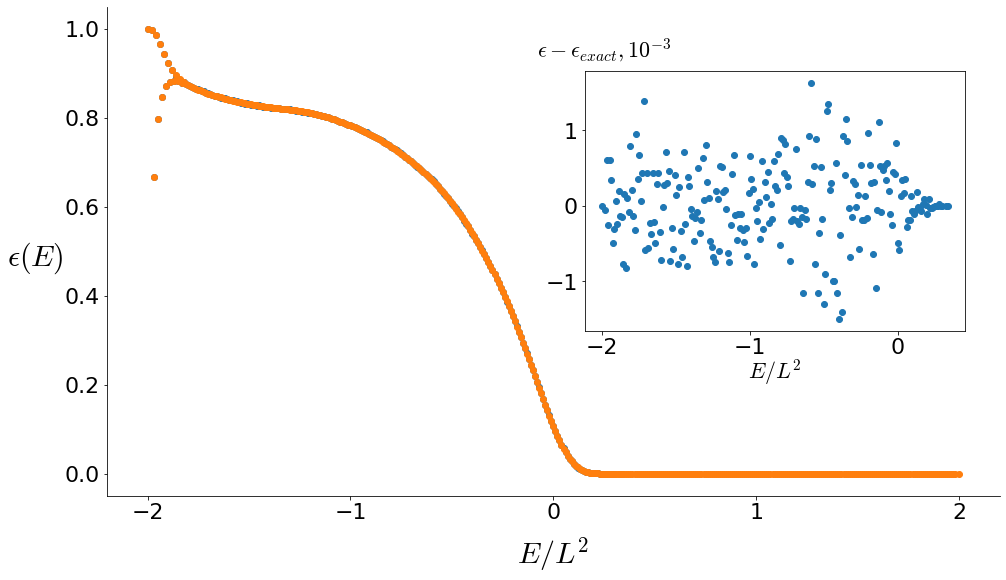}
		\caption{Culling factor $\epsilon(E)$ of a two-dimensional Ising model with linear lattice size $L=40$. Inset: absolute difference between the calculated and exact culling factor, multiplied by a factor of 1000.}
	  \label{fig:culling}
\end{figure}

Figure~\ref{fig:culling} shows the culling factor calculated using the microcanonical population annealing algorithm, and the inset shows the absolute difference between the calculated culling factor and the exact one calculated using Beale's approach~\cite{Beale-1996}. Note that the difference shown in the inset  is multiplied by the factor 1000, and this difference does not exceed $10^{-3}$, and is not visible at the scale of the main figure. 

\begin{figure}[h!]
\includegraphics[width=\linewidth]{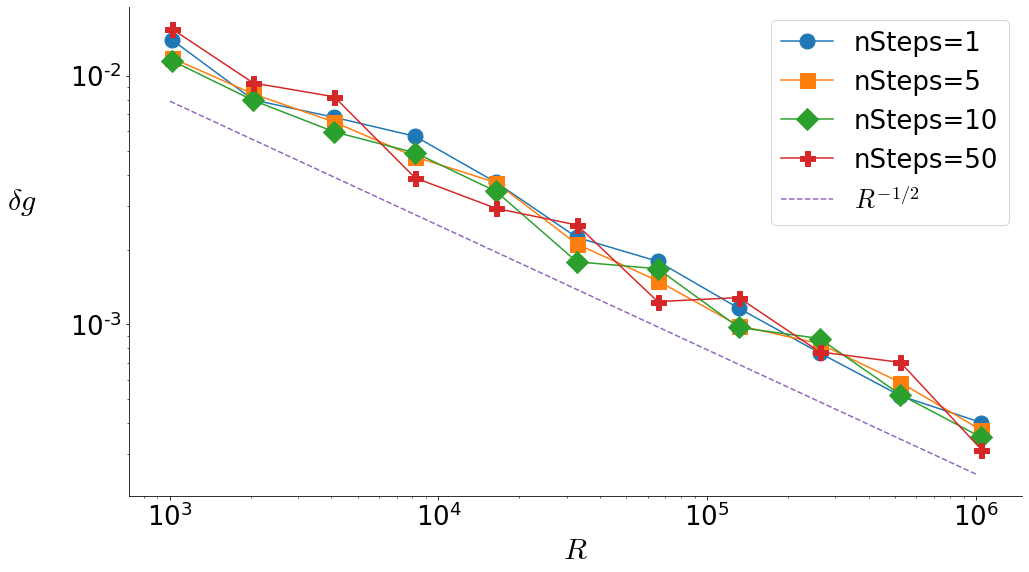}
		\caption{Variation of the relative DoS error $\delta g$ on the number of replicas $R$. The dashed line shows the slop proportional to $1/R^{1/2}$. 2D square lattice Ising model with $L{=}20$.}
	  \label{fig:R-dependence}
\end{figure}

We found no significant effect of the number of MCMC steps $n_s(E)$ on the results, comparing DoS calculated at values of  $n_s(E)$ from 1 to 50. In contrast, the accuracy of DoS strongly depends on the number of replicas $R$. Figure~\ref{fig:R-dependence} shows the variation of the relative DoS error, calculated as the relative difference of the DoS estimate from the MCPA simulation $g_{MCPA}(E)$ from the exact $g(E)$ calculated with Beale's solution~\cite{Beale-1996} $g(E)$
$$\delta g = \sum_{\{E\}}|g_{MCPA}(E)/g(E)-1| $$ as the sum of modulus of the relative differences at each energy level $E$, normalized by the number of energy levels. Remarkably, $\delta g$ behaves like $R^{-1/2}$, reflecting the applicability of the central limit theorem for DoS estimation using MCPA.

We further validated the algorithm by comparing our approach with Wang-Landau simulations using Potts models with 10 and 20 components exhibiting a first-order phase transition, and the results matched well~\cite{Fadeeva-00}. 

The implementation of the Blume-Capel model algorithm has significant differences from the Ising and Potts models.   In general, energy levels are not integers and care must be taken when handling them in the algorithm.
The ceiling or floor goes to the next energy level, and the energy gap depends on the fractional value of $D$. Indeed, $D$ is a finite decimal number, and the last digit $a$ in $D$ is represented as $a10^{-n}$ (e.g., $D{=}1.966$ with $a{=}6$ and $n{=} $3). Since $E{=} \mathrm{integer\; number} {+} D\; \mathrm{integer \; number}$, the minimum energy step is $dD {=} 10^{-n}$ in the general case. Therefore, the ceiling/floor modeling takes orders of magnitude longer as $n$ increases. For example, the number of levels for systems of size $L{=}32$ varies from 4085 at $D{=}0$ to 40589 at $D{=}1.9$ to 383528 at $D{=}1.966$.

 As a result,  we implement automatic detection of energy levels  in the pool and change the ceiling/floor value to the next energy level instead of changing it by $dD$. To avoid the error in comparing floating point numbers, we calculate the energies multiplied by a factor of $10^n$ and use integer arithmetic for them. This is a technical trick, but a very important one, and it should be noted that in the paper we give the energies in $J$ units, so the $10^n$ factor drops out everywhere.

\section{\label{sec:analysis} Simulation results}

\begin{figure}
	\includegraphics[width=\linewidth]{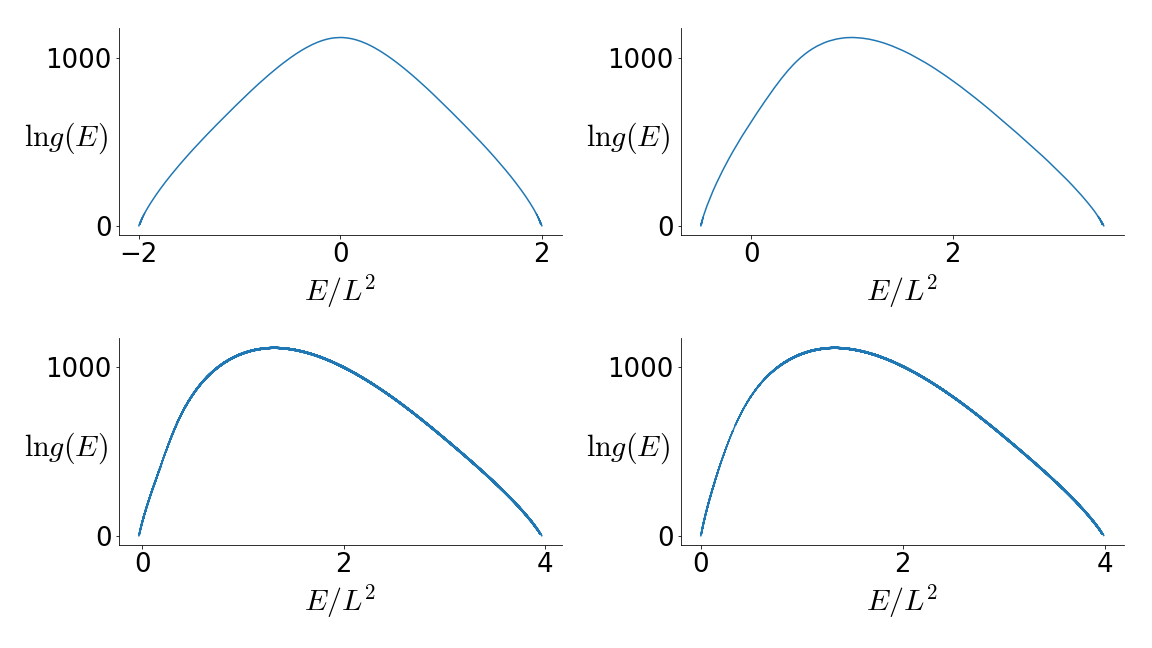}
	\caption{The entropy $S(E)$ is calculated for the ratios $\Delta{/}J{=}0$ and 1.5 in the top row and $\Delta{/}J{=}1.966$ and 1.99 in the bottom row. The number of replicas is $2^{17}$ and the system size is $L=64$.}
\label{fig:entropy}
\end{figure}

 Figure~\ref{fig:entropy} shows examples of entropy estimates for lattices with linear size $L{=}64$ and for several values of the anisotropy parameter $\Delta/J$. It actually displays $\log g(E)$, where $g(E)$ is the density of states. For $\Delta/J{=}0$ it is a symmetric function, as expected, and with increasing values of $\Delta/J$ the maximum shifts to lower energies.

\subsection{Specific heat analysis}

The specific heat is calculated using the expressions~(\ref{eq:E}-\ref{eq:C}). We analyze the finite-size scaling of the height of the specific heat maximum $C_{max}$ and its position $T^C_L$. It is well known~\footnote{The most recent and detailed analysis is provided by Butera and Pernici in the article~\cite{Butera-2018}. } that the phase diagram of the Blume-Capel model, Expr.~(\ref{eq:H}), in the $(T-\Delta)$ plane consists of the lines of the second order and first order phase transitions terminated at the tricritical point. Thus, there are three classes of finite-size behaviour as well as the crossover behaviour around the tricritical point. 
We summarize some estimates of the phase diagram obtained by various methods, as indicated in the figure caption to Fig.~\ref{fig:phase-diagram}. 

The expected finite-size scaling of the specific heat near the second-order line is in the universality class of the 2D Ising model. Accordingly, the two main terms in the finite-size dependence of the specific heat of the Ising model at the critical point behaves~\cite{FF-1969}
\begin{equation}
C=C_0 \left( \ln L +C_1\right) + ...
\label{eq:c-log}
\end{equation}
and a detailed calculation of the terms can be found in~\cite{FF-1969,Fisher-1967,Salas2001,Kolya2002} for a two-dimensional model on the torus, with $C_0{=}8/\pi(J/T_c)^2{\approx} 0.494$. The logarithmic behaviour is universal, but the coefficient $C_0$ is not universal and depends on the details of the Hamiltonian. The fit to the expression~(\ref{eq:c-log}) is given in the second and third columns in the Table~\ref{table1}. These fit is reasonable up to the value of anisotropy parameter $\Delta{/}J{=}1.95$. 

\begin{figure}
  \centering
  \includegraphics[width=.95\linewidth]{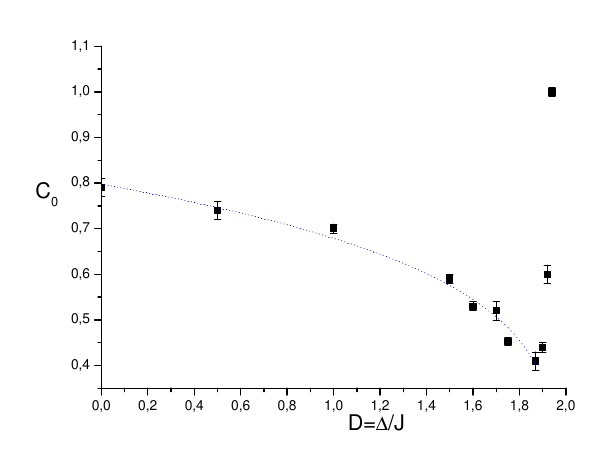}
  \caption{Change of coefficient $C_0$, expression~(\ref{eq:c-log}), with anisotropy parameter $D$. The solid line is the heuristic fitting described in the text. }
  \label{fig:C0-fit}
\end{figure}

It was observed in the paper~\cite{fitC} that the dependence of the coefficient $C_0$ along the critical line on the anisotropy parameter $D$ can be reasonably consistent with the power law  $C_0{\propto} (D-D_{tr})^\omega$ with a reasonable estimate $D_{tr}{=}1.96(1)$. We checked our data against this observation and plot the $C_0$ coefficient from Table~\ref{table1}   along with the fit to the minimum $C_0$ value at $D=1.87$, see Figure~\ref{fig:C0-fit}. Estimating the fit with $\omega{=}0.227(31)$ gives $D_{tr}{=}1.965(53)$, which agrees very well with the widely accepted value of tricritical point estimates~\cite{Butera-2018}. 

\begin{figure}
  \centering
  \includegraphics[width=.95\linewidth]{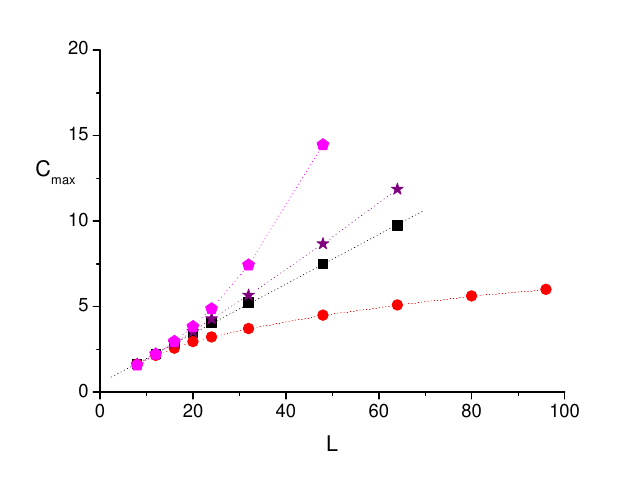}
  \caption{Maximum value of the specific heat of Blume-Cappel model.  Red circles - $\Delta/J{=}1.95$, black squares -  $\Delta/J{=}1.96$,purple stars -  $\Delta/J{=}1.962$, magenta pentagons - $\Delta/J{=}1.966$. }
  \label{fig:BC-C}
\end{figure}

The most typical dependence of the specific heat maximum on the lattice size is shown in Figure~\ref{fig:BC-C}. For the value $0{\le} D{\le}1.95$ corresponding to Table~\ref{table1}, the dependence is logarithmic, as expected for the universality class of the Ising model.

We observe crossover behaviour at a value $\Delta{/}J{=}1.96$, at which the specific heat fits well in a straight line, i.e. with the effective exponent  $\mu_{eff}$ close to 1, as shown in Fig.~\ref{fig:BC-C}. At large values of $\Delta/J$, the divergence of the specific heat grows faster -- the results of fitting the power law to the data of the form
\begin{equation}
C=m_1 L^{\mu_{eff}} + \ldots
\label{eq:c-mueff}
\end{equation}
are given in Table~\ref{table2}. Interestingly, the value of the effective exponent $\mu_{eff}{\approx} 1.6$ for $\Delta/J{=}1.966$ is very close to those expected for the dependence  of the specific heat at the tricritical point ~\cite{Butera-2018}. Indeed, most estimates of the position of the tricritical point coincide with the point ($\Delta/J{=}1.966$, $T_c{=}0.608$), which agrees well with our observations.  Surprisingly, the value of $\mu_{eff}{\approx} 1.6$ was obtained without noticeble corrections to scaling. 

\begin{figure}
  \centering
  \includegraphics[width=.95\linewidth]{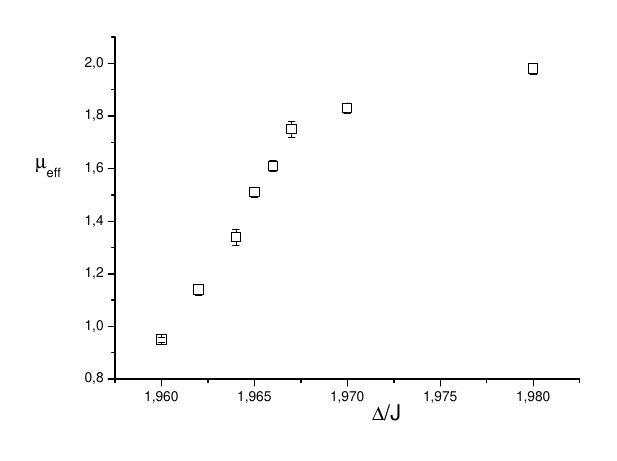}
  \caption{Variation in the effective exponent of the specific heat maximum.}
  \label{fig:mueff}
\end{figure}

Qualitatively, this behavior is similar to the results of the analysis of the phase diagram of the tricritical point based on Landau theory and developed by Bausch~\cite{Bausch-1972}. Figure 4 from his paper illustrates the prediction that there is a very wide crossover region around the first-order critical line, a narrower one around the second-order critical line, and a crossover region that disappears around the tricritical point. We do not claim this explanation, but simply draw the reader's attention to the coincidence, since we found a very strong crossover effect around the first-order critical line. Indeed, the effective exponent $\mu_{eff}$ convergences very slowly to the expected value of 2 for the first-order phase transition , as can be seen from Table~\ref{table2} and Fig.~\ref{fig:mueff}. 

\begin{table}[ht]
\caption{Estimates of the critical temperature $T_c^*$  from the position of the specific heat maximum $C_{max}$, the critical amplitude $C_0$ of the logarithmic behaviour of the specific heat maximum and correction to scaling $C_1$,  and estimate of the critical temperature $T_c^b$ from a Binder cumulant analysis. Ising model universality sector. Linear systems of size $L$ from 16 to 64 were used.}
\label{table1}
\begin{tabular}{|l|l|l|l||l|}
\hline
D/J &   $ T_c^*$ &  $C_0$    &   $C_1$   & $T_c^b$        \\
\hline
 0   &  1.694(4)&   0.79(2) & 0.35(7) & 1.697(1) ) \\ 
 0.5 & 1.5686(4) & 0.74(2) & 0.35(8) & - \\
 1 & 1.401(1) &  0.70(1) & 0.14(7) & 1.400(1) \\
 1.5 & 1.155(1) & 0.59(1) & 0.23(7) & 1.155(2) \\
 1.6 & 1.085(1) & 0.53(1) & 0.40(7) & 1.087(1)  \\
 1.7 & 1.006(1)  &  0.52(2) &  0.33(13) & 1.007(1) \\
 1.75 & 0.961(1) &   0.453(8) & 0.70(7) & 0.9587(4) \\
 1.87 & 0.8203(8) &  0.41(2) & 1.26(19) & 0.8155(5) \\
 1.9 & 0.7731(7) & 0.44(1) & 1.29(14) & 0.7700(7) \\
 1.92 & 0.7338(6) &  0.60(2) & 0.52(10) & 0.731(3)  \\
 1.94 & 0.6891(4) &  1.00(1) & -0.45(3) &  0.688(1)  \\
 1.95 & 0.6613(4) &  1.79(6) & -1.30(7) &  0.6593(4) \\ 
 \hline
\end{tabular}
\end{table}

 \begin{table}[ht]
\caption{Estimates of the critical temperature $T_c^*$  from the position of the specific heat maximum $C_{max}$, the critical amplitude $m_1$ of the power law behavior of the specific heat maximum and the effective critical amplitude $\mu_{eff}$,  and an estimate of the critical temperature $T_c^b$ from Binder's cumulant analysis.}
\label{table2}
\begin{tabular}{|l|l|l|l|l|}
\hline
D/J &  $T_c^*$    &  $m_1$  &$\mu_{eff}$  & $T_c^b$      \\
\hline
1.96 & 0.6321(4)  & 0.18(1) & 0.95(1) & 0.6281(3)\\
1.962  & 0.6219(2) & 0.098(9) & 1.14(2) & 0.6207(5)\\
 1.964 & 0.6156(1) & 0.053(6) & 1.34(3) & 0.6145(2)\\
 1.965 & 0.6121(5) & 0.029(2) & 1.51(2) & 0.6110(1)\\
 {\bf 1.966} & {\bf 0.6089(2)} & 0.020(2) & {\bf 1.61(2)} & 0.6077(1) \\
 1.967 & 0.6057(1) & 0.011(3) & 1.78(6) &  - \\
 1.97   & 0.601(1)     & 0.0110(9) & 1.83(2) & 0.5927(1)  \\
 {\bf 1.98}   & 0.549(1)    & 0.0062(5) & {\bf 1.98(2)} & - \\
 \hline
\end{tabular}
\end{table}

\subsection{Estimation of critical temperature}

\begin{figure}
  \centering
  \includegraphics[width=.95\linewidth]{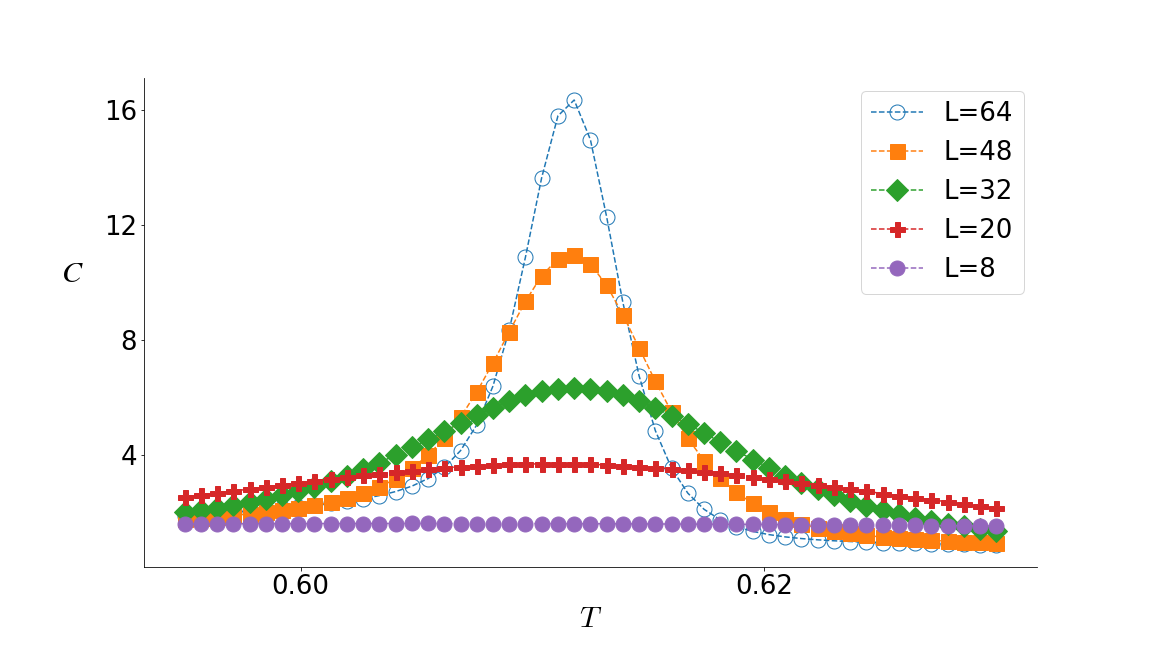}
    \caption{Specific heat in the critical region. $\Delta/J{=}1.965$.}
  \label{fig:spec-heat}
\end{figure}

\begin{figure}
  \centering
  \includegraphics[width=.95\linewidth]{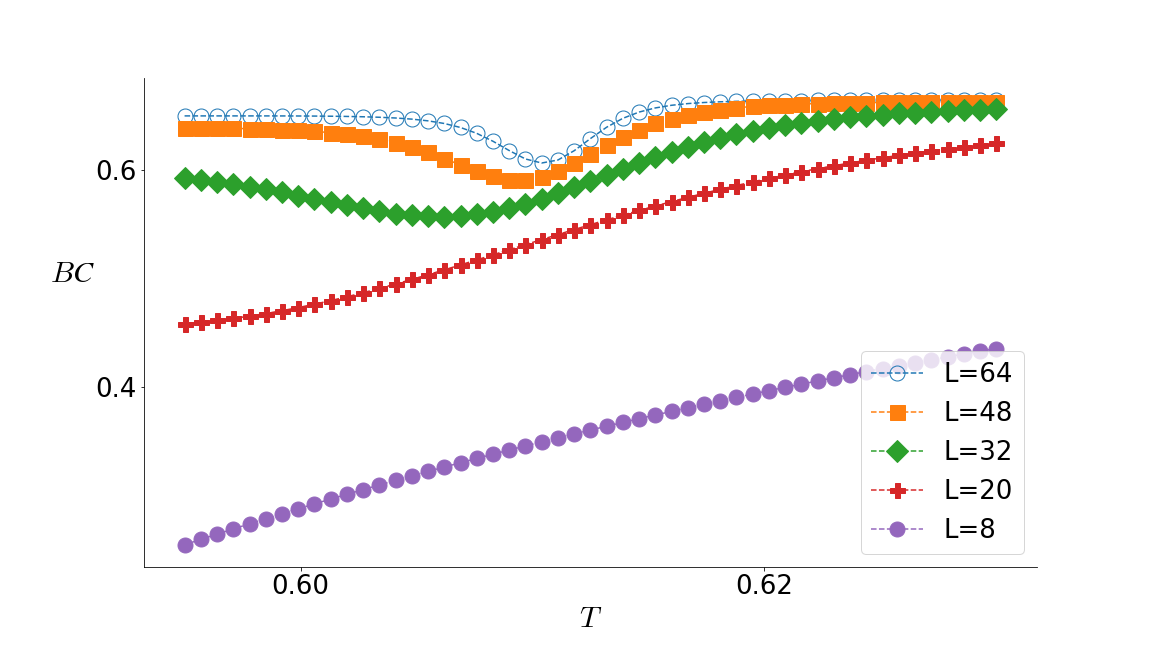}
  \caption{Binder cumulant in the critical region. $\Delta/J{=}1.965$.}
  \label{fig:binder}
\end{figure}

We estimate the critical point from the position of the maximum of the specific heat and the position of the minimum of the Binder cumulant~\cite{Binder-1981,Landau-Binder-book}, that is, from the pseudo-critical temperatures,  which we denote $T_C^*(L)$ and $T_C^b(L)$, respectively, and taking the thermodynamic limit $L\rightarrow \infty$. Figure~\ref{fig:spec-heat} and Figure~\ref{fig:binder} illustrate how specific heat and Binder cumulant behave differently in the critical region depending on the linear lattice size $L$.
Binder cumulant calculated using the second and fourth moments of energy
\begin{eqnarray}
B_E(T) = 1-\frac{\langle E^4(T)\rangle}{3\langle E^2(T)\rangle^2 }.
\label{eq:bind}
\end{eqnarray}

It is known ~\cite{Fisher-1967,FF-1969,Landau-Binder-book}  that the pseudo-critical temperatures shift depends on the correlation length exponent $\nu$ as
\begin{eqnarray}
T_C^*(L)=T_C^* + \frac{a}{L^{1/\nu}}, \;\;   T_C^b(L)=T_C^b + \frac{a}{L^{1/\nu}}
\label{eq:TC}
\end{eqnarray}
where $T_C^*$ and $T_C^b$ are estimates of the critical temperature. The results of the fit~(\ref{eq:TC}) are given in Table~\ref{table1} and Table~\ref{table2} as first and last columns.

We found a difference in the estimates (see Table~\ref{table2}) of the critical temperatures $T_C^*(L)$ and $T_C^b(L)$ from the shift of the heat capacity maximum and the local minimum of the Binder cumulant, a difference growing downward along a first-order line. We attribute this to strong crossover, and more extensive analysis is needed to accurately estimate the first-order critical line.
In contrast, there is no difference in the Table~\ref{table1} estimates of the critical temperature  $T_c^*$ and $T_c^b$ from the displacement of the specific heat maximum  and the local minimum of the Binder cumulant along the line of second-order transitions. Apparently, the influence of the crossover is much weaker in this case.  
It should be noted that the estimates of the critical temperature in the last column of Table~\ref{table2} from the Binder cumulant are closer to Butera and Pernici's estimates of the critical temperature from series expansion~\cite{Butera-2018} than our estimates of the critical temperature from the specific heat.

\begin{figure}
  \centering
  \includegraphics[width=.95\linewidth]{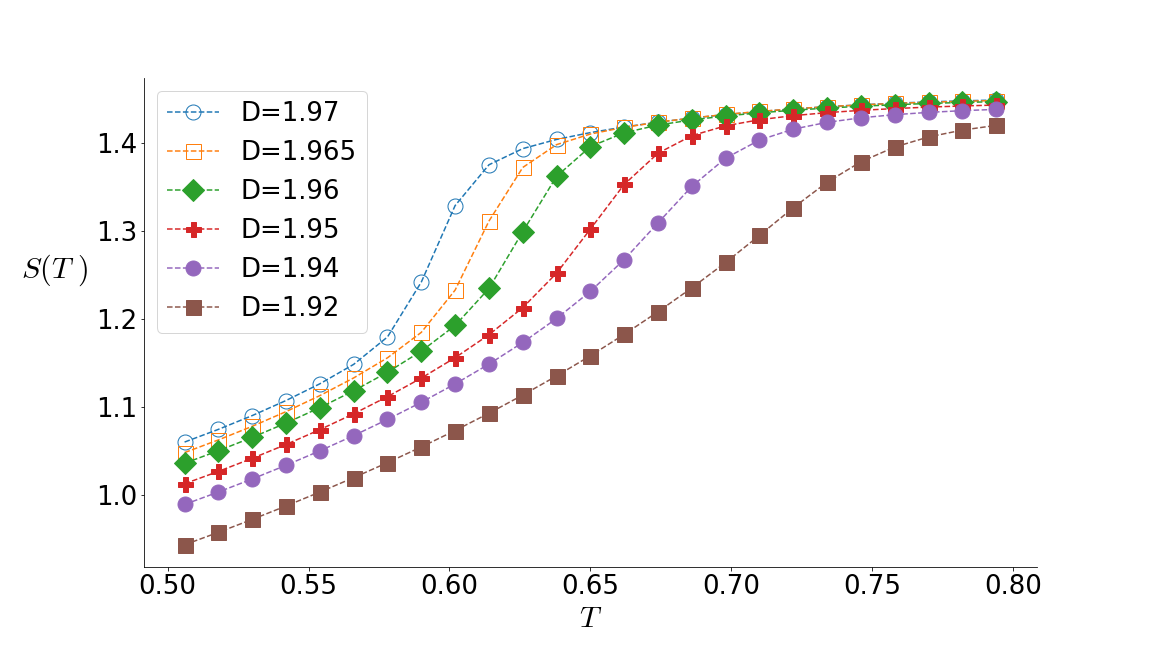}
  \caption{Entropy density for several values of $D{=}\Delta/J$.}
  \label{fig:ED}
\end{figure}

Figure~\ref{fig:ED} shows the entropy change in the critical region for lattice size $L{=}48$ for different values of the disorder parameter $\Delta/J$ in the tricritical point region. This dependence is similar to that presented in Figure~5 of the article~\cite{Jung-2017}, obtained using the transition matrix method. Estimates of the critical temperature from the intersection of the entropy densities for various values of the linear size are consistent with estimates calculated from the specific heat and Binder cumulant behavior.

\section{Discussion}

We used a modified multicanonical population annealing algorithm~\cite{Rose-2019} to analyze the two-dimensional Blume-Capel model. We analyse the finite-size behavior of specific heat and Binder cumulant and observe the evolution from the second-order Ising behavior through the tricritical point to first-order behavior. Our results are in agreement with previous numerical analyses performed by different methods -- Monte Carlo Renormalization Group~\cite{landau1981}, transfer-matrix methods~\cite{Beale-1986,Jung-2017}, real-space renormalization group~\cite{burkhardt1977}, Monte-Carlo method~\cite{XALP98}, Wang-Landau method~\cite{SCP06,Kwak-2015}, microcanonical algorithm~\cite{Zierenberg-2017}, high- and low-temperature expansions~\cite{Butera-2018}, and many others. 

In the Table~\ref{table3} we combined some data from the literature for the critical temperature estimation~\cite{Beale-1986,fitC,Butera-2018} and place in the last column our data for the critical temperature estimation from the position of the specific heat maximum presented also in  the first column of Table~\ref{table1}.
 We cannot compare first-order line estimates because the published data contain four or five digits of the disorder parameter $D$, and our simulation would take a long time if we simulated the system with such a value of $D$. We have to note that the estimations of critical temperature from the Binder cumulant and shown in the last column of Table~\ref{table1} become even more close to the estimations from series expansion published by Butera and Pernici~\cite{Butera-2018} for large values of anisotropy parameter $D/J$.

Our algorithm differs from another class of microcanonical algorithms, such as microcanonical replica exchange algorithm~\cite{Kar-2009}. The main difference is that in MCPA the temperature is not used in the simulation and, therefore, critical slowdown in the usual sense does not occur.  At the same time, the evaporation and condensation of droplets in the vicinity of the first-order phase transition are still determined by the energy barrier associated with surface tension and depend exponentially on the surface length. It is likely that microcanonical simulations is less sensitive to this than  the canonical simulations~\cite{Schierz-2016,Rose-2019}. More research is needed to make this claim more certain.

The accuracy of the data is weakly dependent on the number of MCMC steps, as shown in the Fig.~\ref{fig:R-dependence}. Instead, the accuracy depends on the number of replicas $R$ in the pool and follows a $R^{-1/2}$ behavior. Thus, the main feature of the MPCA is to anneal the population in energy space and estimate the DoS from the averages over a large number of replicas. Therefore, this algorithm is very well suited for massively parallel simulations using a hybrid MPI/GPU approach.
The multicanonical population annealing algorithm is another good approach for modeling critical phenomena.
We found a strong effect of cross-over and finite size near the first-order line, which needs to be explored with more intensive analysis.

\begin{table}[h!]
\centering
\begin{tabular}{|l|l|l|l|l|} 
\hline\hline
$D/J$ & ref.~\cite{Beale-1986} & ref.~\cite{fitC} & ref.~\cite{Butera-2018} & MCPA\\ \hline
 0       & 1.695    & 1.693(3)   & 1.69378(4)  & 1.694(4) \\
 0.5    & 1.567    &  1.564(3)  & 1.5664(1)    & 1.5686(4) \\
 1       & 1.398    &  1.398(2)  & 1.3986(1)    & 1.401(1) \\
 1.5    & 1.150    &  1.151(1)  & 1.1467(1)    & 1.155(1) \\
 1.75  & -           & -                & 0.958(1)      & 0.961(1) \\
 1.87 & 0.800    &  -               & 0.812(1)      & 0.8203(8) \\
 1.9  & -             & 0.769(1)   & 0.766(1)       & 0.7731(7) \\
 1.92  & 0.700   & -                & 0.7289(2)     & 0.7338(6) \\
 1.95  & 0.650   & 0.659(2)    & 0.656(4)       & 0.6613(4) \\
 1.962  & 0.620 &  -              &  -                    & 0.6219(2) \\
 1.966 &  0.610 &  -              &  -                    & 0.6089(2) \\ \hline\hline
\end{tabular}
\caption{Estimates of the critical temperature $T_c/J$  obtained by different methods: transfer-matrix~\cite{Beale-1986}, Wang-Landau~\cite{fitC}, high-energy and low-energy expansions~\cite{Butera-2018}, and microcanonical population annealing (MCPA, our data).}
\label{table3}
\end{table}

\begin{acknowledgments}
We are acknowledge discussions with J. Machta and grateful to N. Rose for providing us with the computational code and for fruitful assistance.

The research was supported by the Russian Science Foundation grant 22-11-00259. 
The simulation was carried out using high-performance computing resources of the National Research University Higher School of Economics. 
\end{acknowledgments}


\end{document}